\documentclass[twocolumn,showpacs,preprintnumbers,amsmath,amssymb, prl]{revtex4}
\usepackage{graphicx}
\usepackage{dcolumn}
\usepackage{bm}

\usepackage{amssymb}
\usepackage{epstopdf}
\newcommand{\be}{\begin{equation}}
\newcommand{\ee}{\end{equation}}
\newcommand{\bea}{\begin{eqnarray}}
\newcommand{\eea}{\end{eqnarray}}

\newcommand{\lp}{\left(}
\newcommand{\rp}{\right)}

\renewcommand{\phi}{\varphi}
\renewcommand{\epsilon}{\varepsilon}

\renewcommand{\Im}{{\rm Im}\,}
\renewcommand{\Re}{{\rm Re}\,}

\begin{document}

\title{
Common-Path Interference and Zener Tunneling in Bilayer Graphene p-n Junctions
}
\author{Rahul Nandkishore and Leonid Levitov}
\affiliation{Department of Physics, Massachusetts Institute of Technology, Cambridge, MA02139}

\begin{abstract}{\bf Interference and tunneling are two signature quantum effects that are often perceived as the yin and yang of quantum mechanics: particle simultaneously propagating along several distinct classical paths versus particle penetrating through a classically inaccessible region via a single least-action path. Here we demonstrate that the Dirac quasiparticles in graphene provide a dramatic departure from this paradigm.
We show that Zener tunneling in gapped bilayer graphene (BLG), which governs transport through p-n heterojunctions, exhibits common-path interference that takes place under the tunnel barrier.  
Due to a symmetry peculiar to the BLG bandstructure, interfering tunneling paths form `conjugate pairs', giving rise to high-contrast oscillations in transmission as a function of the gate-tunable bandgap and other control parameters of the junction. The common-path interference is solely due to forward-propagating waves; in contrast to Fabry-P\'erot-type interference in resonant tunneling structures it does not rely on multiple backscattering.
The oscillations manifest themselves in the junction $I$-$V$ characteristic as N-shaped branches with negative differential conductivity, enabling new high-speed active-circuit devices with architectures which are not available in electronic semiconductor devices. 
}\end{abstract}

\maketitle


Quantum tunneling through two or more barriers that are placed closely together is characterized by transmision which is sharply peaked about certain energies. Such 'resonant tunneling' effect arises because particles can reflect between the barriers and resonate at particular energies, allowing enhanced transmission through the barriers. This resonance phenomenon is similar to that taking place in optical Fabry-P\'erot resonators. Resonant tunneling is particularly desirable in applications since it can give rise to negative differential resistance--current that goes down as voltage goes up--an interesting behavior that can be harnessed to form new devices \cite{Sze,EsakiTsu}.

\begin{figure}[h]
\includegraphics[width =1.02\columnwidth]{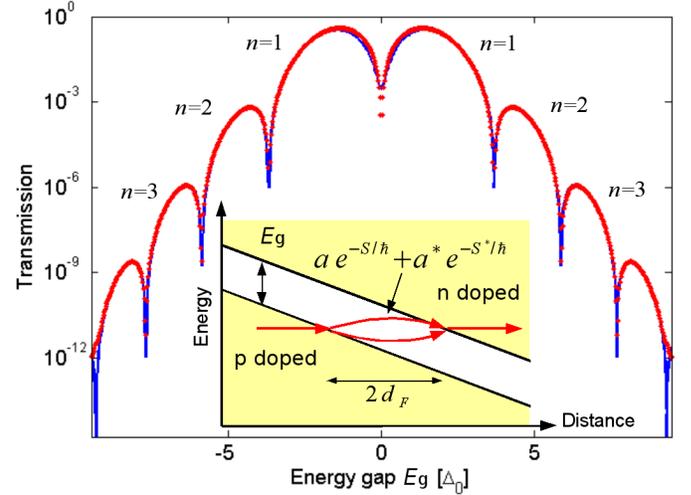}
\caption{Zener tunneling in BLG in the uniform-field model. Interference of two least-action tunneling paths results in oscillations, $n=1,2...$. Shown is transmission at normal incidence, $p_y=0$, as a function of bandgap size, in units $\Delta_0 = \lp (F\hbar)^2/2m\rp^{1/3}$ (semi-log scale). Numerical results (red symbols), obtained by integrating  Eq.(\ref{eq: p repn'}), agree with the WKB result, Eqs.(\ref{eq:SS*}),(\ref{eq:linear-field-results}) (blue curve) in the entire range of  $\Delta$, large and small.
Inset shows schematic setup of p-n junction: the bandgap $E_{\rm g}=2\Delta$, the linear barrier potential $V(x)=-Fx$ [see Eq.(\ref{eq:H psi(x)})], and a pair of interfering tunneling paths. 
}\label{fig: zener oscillations}
\end{figure}

Here we propose an entirely different approach to realize oscillatory tunneling, which involves Zener tunneling of Dirac particles through a p-n junction in gapped BLG \cite{Novoselov,McCannFalko}, a new material with a unique combination of electronic properties, such as the field effect and the possibility to open a bandgap by using external gates \cite{McCann,Oostinga,Crommie}. Interband (Zener) tunneling 
plays a crucial role in materials with several bands of carriers \cite{Zener}. Unlike the conventional tunneling through a potential barrier, which is controlled by the barrier properties, Zener tunneling is governed by an externally applied electric field that produces mixing of states in different bands. Strong enough fields can induce interband transitions from the valence band of p-type material to the conduction band of n-type material, giving rise to tunneling currents. In conventional semiconductors, the tunneling rate is a monotonic function of the applied field $F$ and the bandgap $E_{\rm g}$, given by an exponential $\exp(-\pi m^{1/2}E_{\rm g}^{3/2}/2 F\hbar)$ (here $m$ is an effective mass)\cite{Keldysh,Kane}. In a sharp departure from this behavior, we find that transmission through a p-n junction in BLG {\it oscillates} as a function of the bandgap and external field. The oscillations have 100\% contrast, with transmission vanishing at particular nodal values of control parameters (see Fig.\ref{fig: zener oscillations}).

The interference effects discussed below involve only {\it forward-propagating} waves and a {\it single} barrier, which makes them very diffeent from the Fabry-P\'erot resonances that arise from interference of waves undergoing multiple reflection between several barriers.
The origin of the oscillatory behavior can be elucidated by a semiclassical analysis of the dynamics in the barrier region. In contrast to the standard case of tunneling through a one-dimensional barrier, where a unique saddle-point trajectory in a classically forbidden region is found for each energy, here we obtain multiple trajectories. Further, the trajectories form pairs with {\it complex conjugate} WKB action values $S$ and $S^*$. Such pairs exhibit {\it under-barrier interference}, giving rise to an oscillatory transmission
\be\label{eq:SS*}
T=\left| a e^{-\frac1{\hbar}S}+a^* e^{-\frac1{\hbar}S^*}\right|^2
=4|a|^2 e^{-\frac2{\hbar}S'}\cos^2\lp \frac1{\hbar}S''+\phi\rp
\ee
where $S=S'+iS''$. Both $S'$ and $S''$ are monotonic functions of the bandgap and field strength (see Eq.(\ref{eq:linear-field-results})). These oscillations will manifest themselves through negative differential conductivity in the $I$-$V$ characteristic. 

Relativistic-like behavior of carriers in graphene leads to many interesting transport phenomena\cite{Katsnelson,Cheianov,Martin,Beenakker,CastroNeto}.
However, the oscillatory tunneling discussed here has not been anticipated by theory and is entirely different from Zener tunneling 
that governs transport in p-n junctions in semiconducting carbon nanotubes \cite{Appenzeller,Bosnick} and single layer graphene\cite{Vandecasteele}. Theory of these systems \cite{Andreev,Jena} yields exponential dependences that match closely those of Refs.\cite{Keldysh,Kane}.
Similar exponential dependence arises in the theory of p-n junctions in gated gapless graphene sheets \cite{CheianovFalko}, with a momentum component along the p-n interface playing the role of a bandgap.

The oscillatory tunneling in BLG opens door for designing new device arcitectures. Because the negative $dI/dV$ arises solely due to single-particle effects, it is completely insensitive to the behavior in the doped region. This represents a distinct advantage compared to resonant-tunneling (Esaki) diodes \cite{Esaki} where the effect of temperature on dopants limit thermal stability. Also, the absence of multiple reflection makes the response time potentially much higher than for resonant tunneling in conventional quantum well heterostructures\cite{Sze}.

To clarify the origin of Eq.(\ref{eq:SS*}), we first consider transmission using the WKB formalism. Gapped BLG in the presence of a barrier potential $V(x)$ is described by a $2\times2$ quadratic Dirac Hamiltonian \cite{McCannFalko}
\be\label{eq:H psi(x)}
H=\lp \begin{array}{cc}
\Delta  & \frac{1}{2m}(p_x+ip_y)^2 \\
\frac{1}{2m}(p_x-ip_y)^2 & -\Delta  \end{array} \rp +V(x)
,\quad \Delta=\frac{E_{\rm g}}2,
\ee
where $E_{\rm g}$ is the bandgap.
We seek the wavefunction in the barrier region in the form $\psi(x)\propto e^{\frac{i}{\hbar}\int_{x_0}^xp(x')dx'}\chi$, where $\chi$ is a two-component spinor. The $x$ dependence of momentum can be found from the energy integral $E=\pm\lp \lp p^2/2m\rp^2+\Delta^2 \rp^{1/2}+V(x)$. In the barrier region, $-\Delta <V(x)-E<\Delta$, this gives four complex roots
\be\label{eq:4roots}
p_x(x)=\pm \sqrt{-p_y^2\pm 2m i\sqrt{\Delta^2-(V(x)-E)^2}}
,
\ee
where $p_y$ is a conserved $y$ component of momentum. Two of the roots (\ref{eq:4roots}) have $\Im p>0$, while the other two have $\Im p<0$. Positive (negative) $\Im p$ correspond to decaying (growing) exponentials which describe particle propagation to the right and to the left, respectively.

Focusing on the uniform-field model $V(x)=-Fx$ (see Fig.\ref{fig: zener oscillations} inset) and for simplicity setting $p_y=0$, we select from (\ref{eq:4roots}) the right-propagating solutions: $p_\pm(x)=(i\pm 1)m^{1/2}\lp\Delta^2-(Fx-E)^2\rp^{1/4}$. These two solutions give complex conjugate WKB transition amplitudes $e^{-S/\hbar}$, $e^{-S^*/\hbar}$, where
\be\label{eq:linear-field-results}
S,\, S^*=(1\pm i)\alpha m^{1/2}\Delta^{3/2}/F
\ee
with the prefactor expressed through the Euler beta function,
%
$\alpha ={\,\rm B}(\frac12,\frac54)\approx 1.75 $. 

The total transmission amplitude in the WKB approximation is the sum of the transmission amplitudes for the two decaying exponentials.
Combining the contributions of the trajectories $p_\pm(x)$ we can write the WKB wavefunction in the barrier region as a sum $a e^{\frac{i}{\hbar}\int_{x_0}^x p_+(x')dx'}+a^* e^{\frac{i}{\hbar}\int_{x_0}^x p_-(x')dx'}$. Interference between these evanescent solutions produces an oscillatory transmission amplitude
\be\label{eq:ASS*}
A=a e^{-S/\hbar}+a^* e^{-S^*/\hbar}
.
\ee
Since $\Re S=\Re S^*$ and $\Im S=-\Im S^*$, the two contributions to the transmission amplitude are of equal magnitude and differ in phase by $\Delta\theta=2(\frac1{\hbar}\Im S-\phi)$. Here, $\phi={\rm arg}(a)$ is a phase offset between the two decaying exponentials which can in principle be obtained by matching solutions at the classical turning points, but in practice is more easily obtained through a numerical procedure, which gives $\phi\approx \pi/2$ (see below). 

For certain nodal values of the field strength $F$ and the gap $\Delta$ the interference is destructive, and the transmission probability vanishes. Plugging the values (\ref{eq:linear-field-results}) in Eq.(\ref{eq:ASS*}), we see that the transmission probability $T=|A|^2$ oscillates, going through nodes when $\alpha m^{1/2}\Delta^{3/2}/F\hbar$ is an integer multiple of $\pi$. This gives the nodal values
\be\label{eq:nodal_Delta}
\Delta_n=(\pi n/\alpha)^{2/3}(F^2\hbar^2/m)^{1/3}
,\quad
n=1,2,3...
\ee
that match closely the nodes found numerically, Fig.\ref{fig: zener oscillations}.

The oscillations in transmission, being a general feature deriving from interference, are a robust and generic phenomenon. In particular, the oscillations do not require a linear potential in the barrier region, and the WKB analysis may be straightforwardly generalized to an arbitrary potential profile $V(x)$. Weak perturbations to the BLG dispersion also can be easily incorporated in the above analysis and shown not to matter as long as the perturbation strength is weak compared to the gap $\Delta$. For example, the trigonal warping interaction can affect the dispersion within few ${\rm meV}$ of the Dirac point\cite{McCannFalko}, thus its effect will be small in systems with gate-induced gap that can reach a few hundred ${\rm meV}$ \cite{Crommie}.

Another requirement on experimental systems in which the interference phenomena described above can be realized is that of ballistic transport in the p-n junction region. Recent observation of Fabry-P\'erot (FP) oscillations in graphene p-n-p junctions \cite{YoungKim} provides a clear signature of ballistic transport in this system. The oscillation could be seen for the p-n interface separation of up to $60\,{\rm nm}$, which sets a lower bound on the mean free path in the presence of a top gate. For a rough estimate, writing $F=U/L$ with $U$ a gate-induced potential difference across a p-n junction and $L$ the junction width (see Fig.\ref{fig: ivcurve} inset), from Eq.(\ref{eq:nodal_Delta}) we predict the number of experimentally accessible nodes 
\be
n\approx \frac{\alpha}{\pi} \frac{m^{1/2}\Delta^{3/2}}{F\hbar}
=\frac{\alpha}{\pi\sqrt{2}}\frac{\Delta}{eU}\frac{L}{\ell_\Delta}
,\quad
\ell_\Delta=\frac{\hbar}{\sqrt{2m\Delta}}
.
\ee
For $\Delta=100\,{\rm meV}$, and using the effective mass in BLG $m=0.033\,m_0$,
we estimate the characteristic lengthscale $\ell_\Delta \approx 3.18\,{\rm nm}$.
Taking $eU = 4\Delta$ and $L= 60\,{\rm nm}$, we arrive at $n\approx 4$, which indicates that oscillatory Zener tunneling is well within reach of current experiments.

We now explain the origin of the oscillations from a different perspective, by mapping the transmission across the p-n junction to evolution of a two level system which is swept through an avoided level crossing. This alternative formalism is specialized for the uniform-field model, and thus is less general than the WKB method. However, it provides intuition and affords an independent check on the WKB results by allowing us to numerically evaluate the transmission probability without any undetermined phase offsets. 

The key to this alternative formulation is an observation that, for the uniform-field model $V(x)=-Fx$, the problem greatly simplifies in the momentum representation. Indeed, since $x=i\hbar\partial_{p_x}$, the Schrodinger equation with the Hamiltonian (\ref{eq:H psi(x)}) turns into a first order differential equation 
\begin{equation}\label{eq: p repn'}
i \hbar F  \frac{\partial \psi}{\partial p_x}= \left(\frac{p_x^2 - p_y^2}{2m} \sigma_1 + \frac{2 p_x p_y}{2m} \sigma_2 + \Delta \sigma_3 \right)  \psi 
,
\end{equation}
where the $\sigma_i$ are the Pauli matrices in sublattice space. This equation is identical to the time-dependent Schrodinger equation for a spin-$1/2$ wavefunction with $p_x$ playing the role of time.

\begin{figure}[h]
\includegraphics[width =1.02\columnwidth]{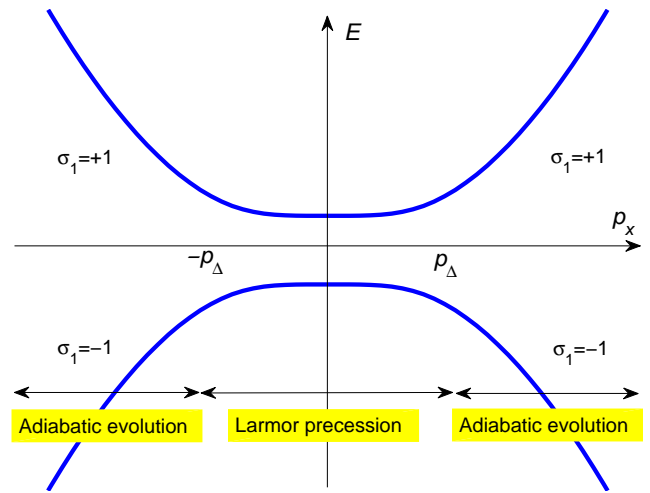}
\caption{Evolution of a two-level system slowly driven through an avoided level crossing, Eq.(\ref{eq: p repn'}). Non-adiabatic transitions between different levels, which correspond to Zener tunneling, take place in the Larmor precession region $-p_\Delta \lesssim p\lesssim p_\Delta $, where $p_\Delta =\sqrt{2m\Delta}$. Shown are adiabatic energy levels of the Hamiltonian, Eq.(\ref{eq: p repn'}) (blue line) and schematic partition into 
regions of adiabatic evolution and Larmor precession.}
\label{fig: anticrossing}
\end{figure}

There is a simple relation between the `time evolution' governed by Eq.(\ref{eq: p repn'}) and interband transitions induced by Zener tunneling \cite{KaneBlount}. Asymptotically, at $p_x \rightarrow \pm \infty$, the eigenstates of the Hamiltonian are also eigenstates of $\sigma_1$, having energies $E_{\sigma_1 = \pm 1}=\mp p_x^2/2m$. As we tune $p_x$ from $-\infty$ to $+\infty$, the system is swept through an avoided level crossing, as illustrated in Fig.\ref{fig: anticrossing}. Interband transitions are described by the process in which a state that started off in the $\sigma_1 = - 1$ eigenstate at $p_x = -\infty$ will evolve into the $\sigma_1 = +1$ eigenstate at $p_x= + \infty$. The evolution is near-adiabatic at small $F$, with Zener tunneling described as (non-adiabatic) transitions across the gap. 

In this framework, the oscillations in transmission can be understood in a simple and intuitive way by noting that the Heisenberg evolution of momentum $p_x(t)$ corresponds to sweeping through the avoided crossing at a constant speed, $dp_x/dt=F$. Comparing different terms in Eq.(\ref{eq: p repn'}), we conclude that transitions may only happen in the region $-p_\Delta \lesssim p_x \lesssim p_\Delta $, where $p_\Delta  =\sqrt{2m\Delta}$ (see Fig.\ref{fig: anticrossing}), whereas outside this region the evolution is adiabatic (here we set $p_y=0$ for simplicity). In the transition region the dominant term in the Hamiltonian is $\Delta\sigma_3$. Spin rotation caused by this term can be described as Larmor precession about the $z$ axis by an angle $\delta \theta=(\Delta/\hbar F)p_\Delta $. Periodic modulation of the transition rate of the form $\cos\delta\theta$, resulting from Larmor precession, leads to an estimate of the oscillation period that agrees with the WKB result, Eqs.(\ref{eq:SS*}),(\ref{eq:linear-field-results}). 

The momentum-sweep analysis helps to understand the dramatic difference between transmission in bilayer junctions and single layer junctions. The latter problem can be mapped \cite{Levitov} to a canonical Landau Zener problem of a linear sweep through an avoided level crossing, for which transmission is a monotonic function of control parameters exhibiting no oscillations. This is in agreement with the theory of p-n junctions in single-layer graphene \cite{CheianovFalko}. 

We now place this discussion on a firm quantitative ground by calculating the transition probability numerically. We solve the differential equation, Eq.(\ref{eq: p repn'}), in a suitably chosen interval $p_{\rm min} <p_x<p_{\rm max}$, taking as the initial state at $p_x=p_{\rm min}$ the adiabatic ground state. From the numerical solution we determine the probability to evolve into the excited state at $p_x =p_{\rm max}$. The transmission probability, obtained in this manner for $p_y = 0$ and $p_{\rm max\, (min)}=\pm 22p_\Delta $, is shown in Fig.\ref{fig: zener oscillations}. The results are compared with the prediction of the WKB approach, Eq.(\ref{eq:SS*}), treating the prefactor $|a|^2$ and the phase $\phi$ as fitting parameters. As illustrated in Fig.\ref{fig: zener oscillations}, excellent agreement is found for the values $\phi=1.6$ and $|a|^2=0.78$ (which are tantalizingly close to $\pi/2$ and $\pi/4$), indicating that the WKB analysis provides reliable results.

Integrating Eq.(\ref{eq: p repn'}) at finite $p_y$ we find that the transmission oscillates and vanishes at nodal values of $\Delta$ in pretty much the same way as for zero $p_y$. Comparing to the WKB analysis, which continues to apply at finite $p_y$, we find that the WKB phase offset $\phi(p_y)$ varies only weakly with $p_y$. Using this numerical procedure, we may also straightforwardly take into account trigonal warping. Apart from a weak washing out of the nodes, we find no significant effect on the oscillations of transmission provided the trigonal warping energy scale is less than the gap size.

\begin{figure}[h]
\includegraphics[width = \columnwidth]{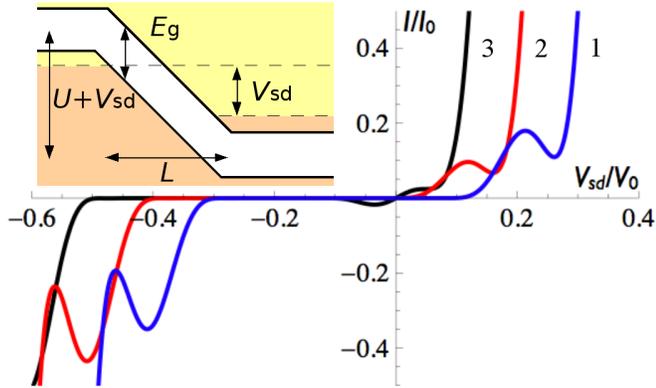}
\caption{
The $I$-$V$ characteristic of a BLG p-n junction combines features of Esaki diode (N-shaped branches with negative differential conductivity) and Zener diode (a breakdown-type behavior). 
Thus a single p-n junction can serve as an active circuit element with multiple functionality. Valleys in the $I$-$V$ dependence correspond to $n=1$ node of the oscillations in transmission in Fig.1.  Shown is the $I$-$V$ dependence given by Eq.(\ref{eq: ivcurve}) for parameter values: $U/V_0=0.1,\,0.2,\,0.3$ (curves 1, 2 and 3, respectively). Units are $V_0 = \Delta (L/\ell_\Delta)$ and $I_0 = 10^{-4} N\frac{e^2}{h} (W/2 \pi\ell_{\Delta})V_0$, where $W$ is the lateral width of the junction and $N=4$ is the spin/valley degeneracy in BLG. Inset shows junction schematic, with $U$ the built-in potential induced by doping or by gates, and $E_{\rm g}=2\Delta$ the bandgap. 
\label{fig: ivcurve}}
\end{figure}

Next, we proceed to show that the oscillatory tunneling reveals itself through distinct features in the $I$-$V$ characteristic. The net tunneling current can be expressed, according to the Landauer formula \cite{Beenakker}, as a sum of contributions of all conducting channels multiplied by energy distribution in reservoirs, giving
\bea\label{eq:I-V_general}
&& I=\frac{e}{h}\int_{-\infty}^\infty dE \lp n_{E-\frac12 eV_{sd}}-n_{E+\frac12 eV_{sd}}\rp \,\mathbf{T}(F)
,
\\\label{eq:netT}
&& \mathbf{T}(F)=\frac{NW}{2\pi\hbar}\int_{-\infty}^\infty dp_y T_{p_y,E}(F)
,
\eea
where $W$ is the total length of the p-n interface, and the factor $N=4$ is spin/valley degeneracy in BLG. Here, accounting for the fact that transmission is dominated by small values of $p_y$ (see below), we treat the occupation numbers as $p_y$ independent and factor out the quantity $\mathbf{T}(F)$, the net transmission integrated over $p_y$.

Continuing to work with the uniform-field model, we treat transmission as energy independent and incorporate the source-drain voltage in the effective barrier potential via $F=F_0+eV_{sd}/L$ (see Fig.3 inset).
Integrating over energies we have
\be \label{eq: ivcurve}
I=\frac{e^2}{h}V_{sd} \mathbf{T}(F_0+eV_{sd}/L)
,\quad
F_0=eU/L
.
\ee
The dependence of transmission $T_{p_y}$ on $p_y$ may be found from Eq.(\ref{eq:SS*}) with $S(p_y)$ and $S^*(p_y)$ evaluated using Eq.(\ref{eq:4roots}). Since the transmission is exponentially small in the barrier width, and the width of the barrier region grows monotonically with $p_y^2$, the net transmission $\mathbf{T}$ is dominated by small values of $p_y$. Hence, we may approximate $S$ and $S^*$ as
\be \label{eq:SS*py^2}
S,\,S^*=i^{\pm 1/2}\alpha \frac{\Delta p_\Delta}{F}+ i^{\mp 1/2} \frac{\tilde\alpha \Delta}{2Fp_\Delta}p_y^2+O(p_y^4)
,
\ee
where $\tilde\alpha=\sqrt{2}B(\frac34,\frac34)\approx 2.4$. Plugging these expressions in Eq.(\ref{eq:SS*}) and performing Gaussian integration over $p_y$, we find
\be\label{eq:net_T_gaussian}
\mathbf{T}(F)=\frac{NW|a|^2 F^{1/2}}{(\pi\tilde\alpha \Delta \ell_\Delta)^{1/2}} e^{-\frac{2}{\hbar}S'}\lp 2^{1/4}+\cos\lp \frac{2}{\hbar}S''+\tilde\phi\rp\rp
,
\ee
$\tilde\phi = 2\phi-\frac{\pi}{8}$, where $S'$ and $S''$ are given by Eq.(\ref{eq:linear-field-results}). Based on numerical results, we ignored the $p_y$ dependence of the phase offset $\phi$ in Eq.(\ref{eq:SS*}). Interestingly, the resulting $I$-$V$ curve, Eq.(\ref{eq: ivcurve}), exhibits {\it negative differential conductivity.}

A more accurate result for the net transmission $\mathbf{T}$ can be obtained by numerical integration of the exact WKB transmission over momenta (see Appendix).
In that, the full  dependence of $S$ and $S^*$ on $p_y$ is retained, and also the contribution of the classically forbidden regions $\Delta<|Fx-E|<\sqrt{\Delta^2+(p_y^2/2m)^2}$ is included, which is of subleading order in $p_y^2$. 

The resulting $I$-$V$ dependence is shown in Fig.\ref{fig: ivcurve} for several values of the `built-in' (gate-induced) potential difference across p-n junction. Strikingly, the $I$-$V$ characteristic combines features of the Zener diode (sharp rise of current above certain breakdown voltage) with N-shaped branches on which the differential conductivity is negative, resembling the resonant-tunneling (Esaki) $I$-$V$ characteristic \cite{Esaki}. Unlike the Esaki characteristic, the N-shaped branches occur simultaneously on the forward and reverse parts of the $I$-$V$ dependence. The N-shaped features arise from oscillatory transmission (described by the uniform-field model), a mechanism very different from that leading to negative $dI/dV$ in the Esaki diode. The valleys of current in Fig.\ref{fig: ivcurve} correspond to nodes of transmission ($n=1$ in Fig.1). 

We note that p-n junctions of the type considered here can be realized using a configuration of gates which is already employed in current experiments \cite{Huard,Williams,Oostinga,YoungKim}. A minimal configuration is a dual-gate geometry with a wide back gate and a narrow top gate, such as that employed in the work on FP oscillations \cite{YoungKim}. Charging the two gates with voltages of opposite polarity, a bandgap can be induced under the top gate and, simultaneously, carrier density can be adjusted in the outer region. Applying source-drain bias will produce $V_{\rm sd}$-dominated lateral electric field across the gapped region, corresponding to the regime $U\ll V_0$ where the effect of oscillations is most prominent (see Fig.\ref{fig: ivcurve}). In addition, a built-in field field $U$ can be induced by selective doping or by a third gate. 

In summary, transport in BLG p-n junctions 
is governed by common-path interference under the tunnel barrier. Unlike Fabry-P\'erot interference that stems from multiple reflection between barriers, our interference effect involves only forward-propagating paths and a single barrier. Common-path interference produces nodes in transmission as a function of the gate-tunable bandgap and other control parameters, leading to a complex $I$-$V$ characteristic combining branches with negative differential conductivity.
The single-particle origin of negative $dI/dV$ makes it insensitive to the behavior in the doped regions which limits thermal stability and operation speed of resonant tunneling (Esaki) diodes \cite{Esaki}. The operation speed is further enhanced compared to quantum-well-based devices by the absence of multiple backscattering\cite{Sze}.
We envision that BLG p-n junctions, owing to their multiple functionality and design simplicity, will become part of the future graphene electronics toolkit.

\acknowledgements{We acknowledge useful discussions with N. Gu, C. M. Marcus and M. Rudner, and support from Naval Research Grant N00014-09-1-0724}.


\section{Appendix}
Our goal here is to calculate the net transmission $\mathbf{T}$, a quantity used to evaluate the total current through p-n junction, Eqs.(9),(11) of the main text. For that we evaluate transmission as a function of $p_y$, where $p_y$ is the momentum parallel to the p-n junction. We will then integrate the transmission over $p_y$ to obtain the total integrated transmission through the p-n junction. We calculate this quantity by working in a WKB approximation. 
 
When the WKB method is applied to the BLG p-n junction, the wave-function $\psi(x)$ is written as a sum of plane waves where the (potentially complex) wave-vectors are solutions of the classical equation 
\begin{equation}
\bigg(\frac{\hbar^2(\kappa^2(x) + k_y^2)}{2m}\bigg)^2 + \Delta^2- (V(x)-E)^2 = 0 \label{eq: complexp2}
\end{equation}
where $\kappa(x)$ is the WKB wavevector $x$ component, $E$ is total energy, $\hbar k_y = p_y$ is the momentum parallel to the p-n junction, $2\Delta$ is the bandgap and $V(x)$ is the gate potential. We are working in the uniform-field model $V(x)=-Fx$ and, without loss of generality, set $E=0$. The classical equation (\ref{eq: complexp2}) has no real solutions for $\kappa$ in the classically forbidden region $|x|<x_f$, where the turning points $\pm x_f(k_y)$ are determined by the condition $\kappa(\pm x_f) = 0$. From Eq.(\ref{eq: complexp2}) we find
%
\begin{equation}
x_f(k_y) = \frac{\Delta}{F} \sqrt{1+  k_y^4 \ell_\Delta^4} \label{eq: xf2}; \qquad \ell_{\Delta} = \frac{\hbar}{\sqrt{2m\Delta}}
,
\end{equation}
where $\ell_\Delta$ is a lengthscale set by the gap. For $|x|<x_f$, Eq.(\ref{eq: complexp2}) has no real solutions and the wavefunction is entirely evanescent.

In this problem, the forbidden region consists of two distinct parts. For $|x|<  \Delta/F$, Eq.(\ref{eq: complexp2}) has four complex solutions for $\kappa$, given by Eq.(3) of the main text. Of these four complex solutions, two correspond to tunneling from right to left and may be neglected, whereas the other two correspond to tunneling from left to right, with equal decay constants and a relative phase, which interfere when combined together.

At $|x| = \Delta/F$, there is a doubly degenerate pure imaginary solution to Eq.(\ref{eq: complexp2}), which corresponds to tunneling from left to right. In the outer part of the forbidden region $\Delta/F < |x| < x_f$, the expression Eq.(\ref{eq: complexp2}) has four pure imaginary solutions, of which two correspond to tunneling from left to right. However, in this regime, the two tunneling paths are non-degenerate (have different decay constants, $\Re S_1\ne \Re S_{1'}$), and we consider tunneling only along the path with the longer decay length. 

In the semiclassical approximation, the amplitude of tunneling across the entire forbidden region, from $x = - x_f$ to $x = x_f$ is 
\begin{equation}
A(k_y) = \bigg[a \exp\bigg(-\frac{S}{\hbar}\bigg) + a^* \exp\bigg(-\frac{S^*}{\hbar}\bigg)\bigg] \exp\bigg(-\frac{2S_1}{\hbar}\bigg) \label{eq: WKB}
\end{equation}
Here the action $S$ is accumulated in the region $-\Delta/F < x < \Delta/F$. In this region there are two 'conjugate' tunneling paths with actions $S$ and $S^*$, which interfere as discussed in the main text. The action $S_1$ is accumulated in the outer regions $-x_f<x<-\Delta/F$ and $\Delta/F<x<x_f$, where the tunneling paths are non-degenerate. In this region we consider only the tunneling path with longer decay length. The constants $a$ and $a^*$ are parameters that may be found in principle by matching solutions at the classical turning points (but in practice are treated as fitting parameters with values obtained from numerical solution). The WKB actions are given by
\bea\label{eq: action}
&& S=\hbar\int_{-\Delta/F}^{\Delta/F} \frac{dx}{ \ell_\Delta} \sqrt{k_y^2 \ell_\Delta^2 + i \sqrt{1 - (Fx)^2/\Delta^2}} 
,\quad 
\\
&& S_1=\hbar \int_{\Delta/F}^{x_f}  \frac{dx}{\ell_\Delta} \sqrt{k_y^2 \ell_\Delta^2 - \sqrt{(Fx)^2/\Delta^2 -1}}
\eea
where $\ell_\Delta$ and $x_f$ are given by Eq.(\ref{eq: xf2}). Note that the action $S$ is complex (has real and imaginary parts), whereas the action $S_1$ is pure real. The tunneling probability is given by the square of the tunneling amplitude, $T(k_y)=|A(k_y)|^2$, and takes the form
%
\bea \label{eq: transmission}
T(k_y) = && \mathcal{N}  \exp\bigg( - 2\frac{\Delta}{F \ell_\Delta} \Re f(k_y^2 \ell_\Delta^2)\bigg) 
\\ \nonumber
&&\times \cos^2\bigg(\frac{\Delta}{F \ell_\Delta} \Im f(k_y^2 \ell_\Delta^2) + \phi(k_y)\bigg)
,
\\ \label{eq: f}
f(k_y^2) = && 
\int_{-1}^{1} \sqrt{k_y^2\ell_\Delta^2 + i \sqrt{1 - v^2}} dv 
\\ \nonumber
&&+ 2\int_1^{Fx_f/\Delta} (k_y^2\ell_\Delta^2 - \sqrt{v^2-1})^{1/2} dv 
\eea
%
where the first term describes the conribution of the region $|x|\le \Delta/F$, and the last term accounts for contributions of the outer regions $-x_f<x<\Delta/F$ and $\Delta/F<x<x_f$ (see Eq.(\ref{eq: action})). Here $x_f$ and $\ell_\Delta$ are defined by Eq.(\ref{eq: xf2}), and the quantities $\phi(p_y) = 2\arg(a)$ and $\mathcal{N} = 4|a|^2$ are fitting parameters that may be found in principle by matching solutions at the classical turning points. Note that the tunneling probability is oscillatory at each $k_y$, with the oscillations coming from the imaginary part of $f$. The imaginary part of $f$ comes entirely from the region $|x|<\Delta/F$, where the action $S$ is complex, and where there are two degenerate tunneling paths, which interfere. 
 \begin{figure}[h]
 \includegraphics[width = 1.02\columnwidth]{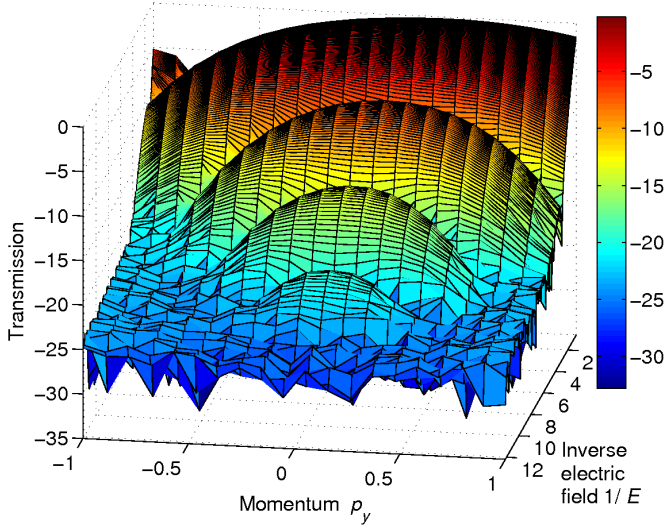}
 \caption{Transmission (semi-log scale) as a function of $p_y$ and electric field. Units: $p_\Delta=\sqrt{2m\Delta}$ and $F_0=\Delta/\ell_\Delta$. \label{fig: pytransmission}} 
 \end{figure}

Fitting to the numerical solution of the BLG Hamiltonian, obtained by integrating time-dependent Schroedinger equation with a $2\times 2$ Hamiltonian given by Eq.(8) of the main text, we obtain $\mathcal{N} \approx 3.2$ and $\phi(k_y = 0) \approx 1.6$ (see Fig.1 in the main text). 

The positions of the nodes, however, depend on $k_y$. There are two main sources of this dependence. First, the period and phase of the oscillations is controled by the imaginary part $\Im f$ dependence on $k_y$.
Second, the phase offset $\phi(k_y)$ can vary with $p_y$, producing additional shift of the nodes. However, since the contribution of $\Im f$ to the net phase is greater than that of $\phi(k_y)$ by a large factor $\Delta/F\ell_\Delta\gg1$ (see Eq.(\ref{eq: transmission})), we expect the node positions variation with $k_y$ to be dominated by $\Im f$.

To compare the two effects, we use numerical solution to find transmission as a function of $p_y$ (see Fig.1). Mapping out the nodes, we find that the phase offset $\phi$ varies only weakly with $k_y$ over the range $k_y^2\ell_\Delta^2<1$ that dominates the integral. The position of the nodes is thus controlled mostly by modulation of the period. 
Hence, we approximate by taking $\phi \approx 1.6$ for all $k_y$. 

\begin{figure}[h]
\begin{center}
\includegraphics[width =  0.8\columnwidth]{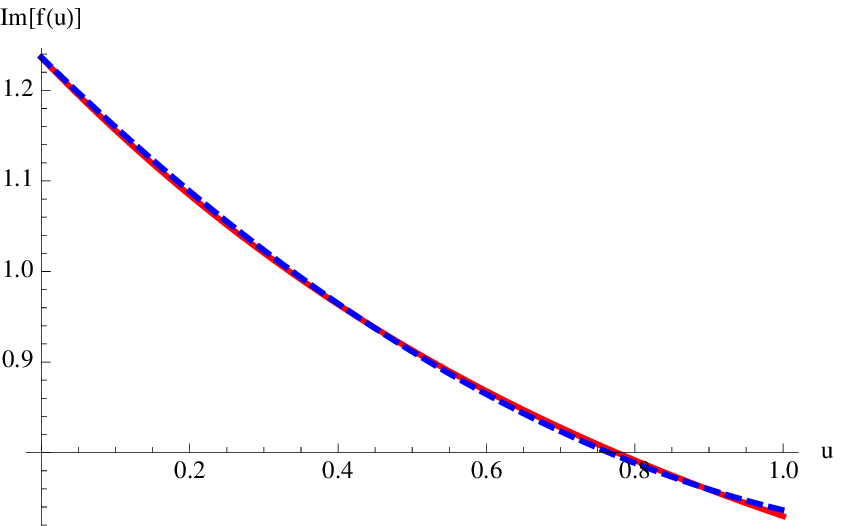}
\includegraphics[width =  0.8\columnwidth]{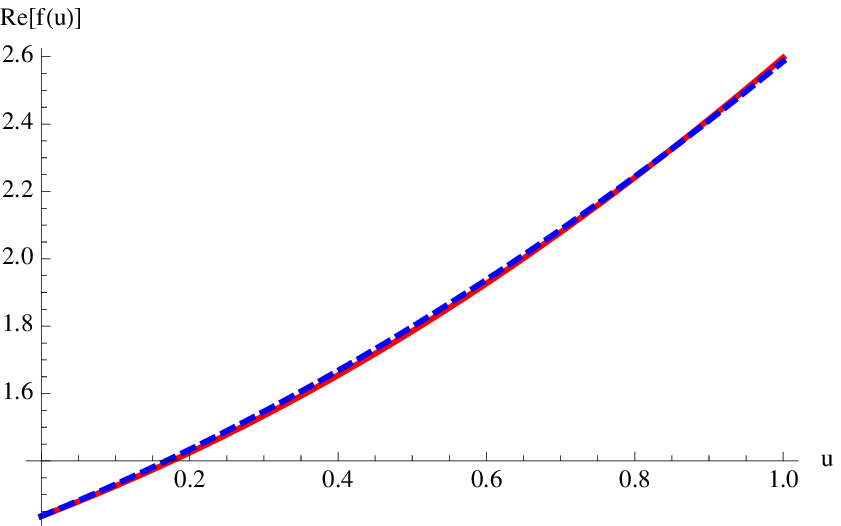}
\end{center}
\caption{Shows fit of function $f(u)$, defined by Eq.(\ref{eq: f}), to the second order polynomial Eq.(\ref{eq: fit}), separately for the real and imaginary part of $f$. The solid red line is the exact curve, the dashed blue line is the polynomial fit. \label{fig: fit}}
\end{figure}

We now wish to calculate
\begin{equation}
\mathbf{T} = N \sum_{k_y} T(k_y) \approx \frac{NW}{2\pi} \int dk_y T(k_y) \label{eq: integratedt}
\end{equation}
where we have assumed that the $pn$ junction has lateral width $W$, and the factor of $N=4$ arises from summing over spins and valleys. We introduce the variable $u = k_y^2 \ell_\Delta^2$. The integral Eq.(\ref{eq: integratedt}) is dominated by small values of $u$. This is illustrated in Fig.\ref{fig: pytransmission}, which plots the transmission as a function of $p_y$ and electric field, as obtained from numerical calculation using the momentum sweep model. It is clear from the figure that transmission is dominated by small values of $p_y$, i.e. $u < 1$. Therefore, we fit $f(u)<1$ in Eq.(\ref{eq: f}) to a polynomial, and obtain
\begin{equation}\label{eq: fit}
f(u) \approx 1.236 + 0.9 u + 0.45 u^2 + i (1.236 - 0.8 u + 0.3 u^2) + O(u^3) 
.
\end{equation}
This second order polynomial fit provides an excellent approximation to $f(u)$, as illustrated graphically  in Fig.\ref{fig: fit}.

We then restrict the integral Eq.(\ref{eq: integratedt}) to $0<u<1$, to obtain
\begin{equation}
\mathbf{T} = \frac{N W}{2 \pi \ell_\Delta} \Phi(\Delta/F\ell_\Delta) \label{eq: totaltrans}
\end{equation}
where the function $\Phi$ is  
\bea\nonumber
\Phi(\Delta/F\ell_\Delta)&&=\int_0^{1} \frac{du}{\sqrt{u}} \exp\bigg( - 2\frac{\Delta}{ F \ell_\Delta}(1.236 + 0.9 u + 0.45 u^2)\bigg) 
\\ \label{eq:Phi}
&& \times \cos^2\bigg(\frac{\Delta}{F \ell_\Delta} (1.236 - 0.8 u+ 0.3 u^2) + 1.6\bigg). 
\eea
 %
Evaluating numerically the integral over $u$ and plugging the result in Eq.(\ref{eq: totaltrans}) gives
the total transmission, summed over $p_y$.
This result can now be used to obtain the $I$-$V$ characteristic of the p-n junction, as discussed in the main text. 
 

\end{document}